\newcommand{\eq}[1]{Eq.~(\ref{#1})}
\newcommand{\fig}[1]{Fig.~\ref{fig:#1}}
\newcommand{\sect}[1]{Section \ref{sect:#1}}
\newcommand{\tab}[1]{Table~\ref{tab:#1}}
\newcommand{\rol}[1]{Ref.~\onlinecite{#1}}

\documentclass[%
reprint,
superscriptaddress,
showpacs,preprintnumbers,
 amsmath,amssymb,
prb,
]{revtex4-1}
\usepackage{gensymb}
\usepackage{graphicx}% Include figure files
\usepackage{dcolumn}% Align table columns on decimal point
\usepackage{bm}% bold math
\usepackage{subfigure}
\usepackage{color}
\usepackage{pstricks}
\usepackage{epstopdf} % .eps figures
\begin{document}

%\preprint{APS/123-QED}

\title{Properties of heavy rare-gases adlayers on graphene substrates}% Force line breaks with \\

\author{Lucas Madeira}
\email{lucas.madeira@asu.edu}
\affiliation{Department of Physics and Astronomy, Arizona State University, 
Tempe, Arizona 85287, USA}

\author{Silvio A. Vitiello} 
\affiliation{Instituto de F\'{i}sica Gleb Wataghin, Universidade Estadual de 
Campinas, UNICAMP, Campinas 13083-859, Brazil}


%\affiliation{
%\affiliation{

\date{\today}% It is always \today, today,

\begin{abstract}
We investigated properties of heavy rare-gases, Ne, Ar, Kr, Xe and Rn, adsorbed 
on graphene substrates using molecular dynamics.
We gathered evidences of 
commensurate solids for Ne and Kr adlayers,
one of them is given by a
typical behavior
of the nearest neighbor distance of the adatoms.
The specific heat and the melting 
temperature were calculated and both indicate continuous melting for all heavy 
noble-gases studied.
We also determined the distance between the
adlayer and the substrate.

%\begin{description}
%\pacs{68.43.-h,64.70.dj}
%use the optional argument of the \verb+\item+ command to give the category of 
%each item. 
%\end{description}
\end{abstract}

\maketitle

%\tableofcontents
\section{Introduction}
\label{sect:intro}

The study of adsorption on solid surfaces began long ago because of its
intrinsic scientific interest and also because of its importance as a mean to
improve the understanding of physical processes at atomic level, which
can be of technological interest. Thin films adsorbed on substrates are
confined to a narrow domain near the surface of the substrate, and they
are very close to a two-dimensional (2D) phase of matter. The subject of
adsorbed layers of a given material is particularly interesting because of
possible bidimensional phases that have no tridimensional analog.
Furthermore, the spatial periodicity of the substrate may cause
interesting effects when the adsorbate lattice ``feels" the symmetry of
the substrate. Usually, the density distribution and other thermodynamical
properties are quite different from what would be expected from a simple 2D
material not adsorbed in a substrate.

Although the adsorption of noble gases on graphite substrates has been
extensively studied from the theoretical and experimental point of view
(for a comprehensive review of the subject the reader is referred to
\rol{bru07} and references therein), the physics of noble gases on
graphene has not received the same attention.  It is our intention to
extend this large body of knowledge by studying properties of
heavy noble gases adsorbed on graphene substrates, a system that presents
remarkable novel properties when compared with graphite. 

Noble gases adsorbed on graphite present interesting properties and we
restrict ourselves to cite and compare some of the most remarkable of them
with our object of study. The Ne adsorbed on graphite exhibits a superlattice
structure \cite{huf76} known as $\sqrt{7}\times \sqrt{7}$, whereas Kr and
Xe form a $\sqrt{3}\times \sqrt{3}$ commensurate lattice \cite{bru97}. For 
Ar on graphite the
order of the melting transition is unsettled, due to the
observation of two peaks in specific heat measurements of the system
\cite{mig84,fle06}. Radon is set apart from the other noble gases because
it is radioactive. Although there is great interest in Rn filters,
which requires knowledge of adsorption behavior on solid surfaces, little
is known about Rn adsorbed on graphite \cite{per08}.

This work is organized as follows. In the next section, we present the
methods employed in our work.
% It begins with a brief description of how it is organized.
\sect{res} contains our results, in subsection A we discuss
the spatial distribution of the adatoms. We have observed evidences of Ne
$\sqrt{7}\times \sqrt{7}$ and Kr $\sqrt{3}\times \sqrt{3}$ commensurate
solids. For the other noble gases, the adlayer is incommensurate with the
graphene substrate for the temperature and densities ranges considered.
The specific heats are calculated in subsection B and the
melting temperatures of the adlayers are determined in subsection
C, both properties present evidences of a continuous melting
for all heavy rare-gases studied. We also estimated two other properties of
the adlayers: the first neighbor distance, subsection D,  and the distance
between the adlayer and the substrate, in subsection E. We note
that it is possible to
relate the behavior of these quantities directly to the specific heat
peaks or the melting transition. It is also noteworthy that the commensurate
adlayers have a typical behavior of the nearest neighbor spacing near the
melting transition. The discussion of the results and conclusions are
presented in Section \ref{sect:disc}.

\section{Methods}
\label{sect:met}

In this section we start by 
briefly introducing the two commensurate structures, $\sqrt{3}\times
\sqrt{3}$ and $\sqrt{7}\times \sqrt{7}$. The adopted
potential interactions are presented in subsection B. In subsection C we give
some details of the simulations.

\subsection{Commensurate structures}
\label{sect:com}

The usual primitive and basis vectors provide a convenient way of expressing the 
periodicity of both substrate and adlayer lattices.
The graphene sheet is a triangular lattice with a basis of two carbon atoms at
\begin{eqnarray}
\textbf{b}_1 = \frac{(\textbf{a}_1+\textbf{a}_2)}{3} = b \ \hat{\textbf{x}}, 
\qquad 
\textbf{b}_2 = 2 \textbf{b}_1 = 2b \ \hat{\textbf{x}},
\end{eqnarray}
where $b$ is the bond length of two carbon atoms assumed to be $1.42$ \AA \ and 
$\textbf{a}_1$ and $\textbf{a}_2$ are primitive vectors of the 2D Bravais 
lattice of the triangular, or hexagonal close-packed structure, 
\begin{eqnarray}
\label{eq:triangular}
\textbf{a}_1 = a \left( \frac{\sqrt{3}}{2} \hat{\textbf{x}} + \frac{1}{2} 
\hat{\textbf{y}} \right), \qquad
 \textbf{a}_2 = a \left( \frac{\sqrt{3}}{2} \hat{\textbf{x}} - \frac{1}{2} 
 \hat{\textbf{y}} \right).
\end{eqnarray}

The so called $\sqrt{3}\times\sqrt{3}R30\degree$ adlayer commensurate structure 
is a triangular lattice of constant $3b$ and axes rotated by 30$\degree$ 
relative to the underlying triangular lattice. %, Fig.~\ref{fig:sqrt3}.
The $\sqrt{3}\times\sqrt{3}R30\degree$ primitive vectors $\textbf{c}_1$ and 
$\textbf{c}_2$ are obtained from \eq{eq:triangular},
\begin{eqnarray}
\label{eq:sqrt3}
\textbf{c}_1 = 3b\hat{\textbf{x}}, \qquad
\textbf{c}_2 = \frac{3b}{2}\hat{\textbf{x}}+
\frac{3b\sqrt{3}}{2}\hat{\textbf{y}},
\end{eqnarray}
with the identification $a=3b$.

\begin{figure}[!htb]
  \centering
  \includegraphics[width=0.5\linewidth]{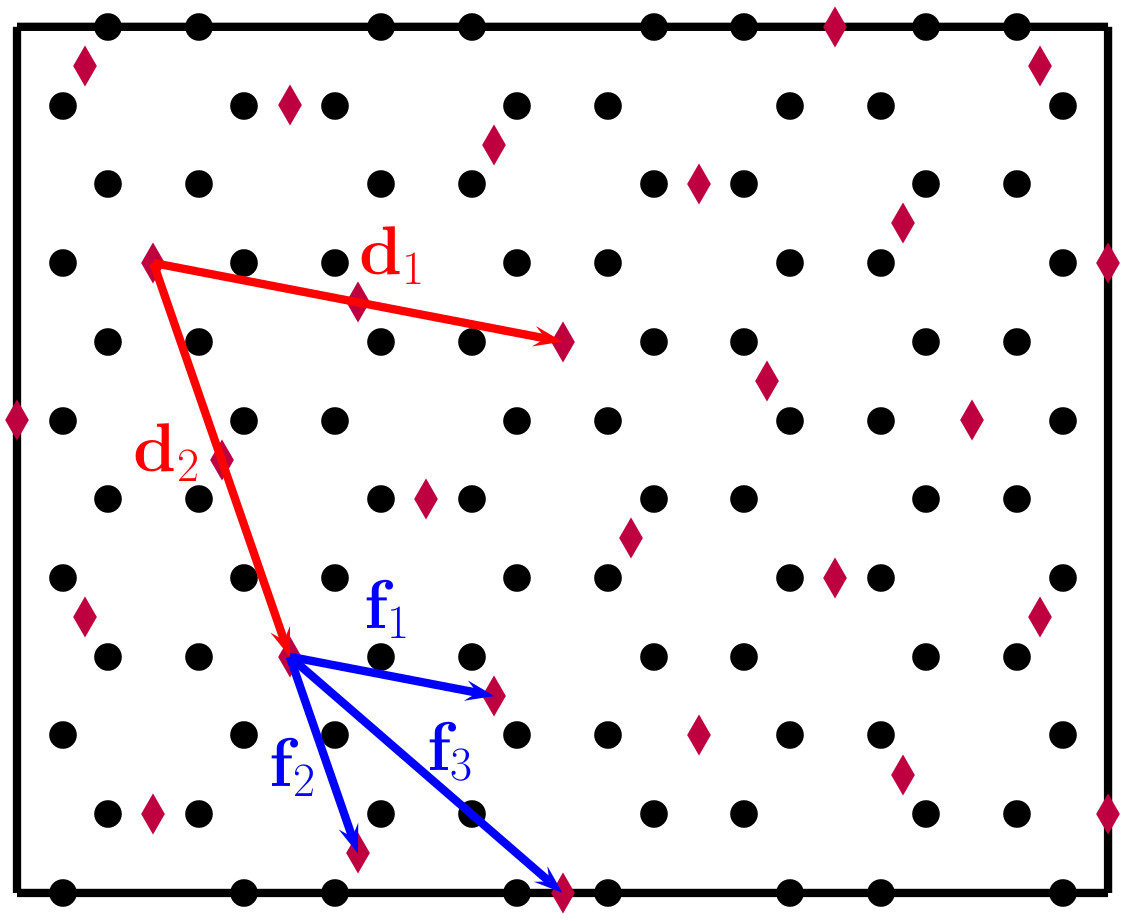}
  \caption{(Color online)
Commensurate $\sqrt{7}\times\sqrt{7}R19.1\degree$ lattice
denoted by diamonds ($\blacklozenge$),
the circles ($\bullet$) stand for
the graphene honeycomb lattice. Vectors $\textbf{d}_i$
and $\textbf{f}_i$
are defined in
\eq{eq:sqrt7_pri}
and \eq{eq:sqrt7_bas}, respectively}
  \label{fig:sqrt7}
\end{figure}

The structure $\sqrt{7}\times\sqrt{7}R19.1\degree$, Fig. \ref{fig:sqrt7},
is a superlattice with primitive vectors
\begin{eqnarray}
\label{eq:sqrt7_pri}
\textbf{d}_1 = \textbf{a}_1+2\textbf{a}_2, \qquad \textbf{d}_2 = 
3\textbf{a}_2-2\textbf{a}_1,
\end{eqnarray}
and four atoms in the basis, one at a superlattice site, the origin,
and three atoms at
\begin{eqnarray}
\label{eq:sqrt7_bas}
\textbf{f}_1 = \frac{\textbf{d}_1}{2}, \qquad
\textbf{f}_2 = \frac{\textbf{d}_2}{2}, \qquad
\textbf{f}_3 = \frac{\textbf{d}_1+\textbf{d}_2}{2}.
\end{eqnarray}
Notice that $|\textbf{d}_1|=|\textbf{d}_2|=a\sqrt{7}$, and the angle
between the vectors $\textbf{d}_1$ and $\textbf{a}_1$ is 19.1$\degree$,
hence the terminology $\sqrt{7}\times\sqrt{7}R19.1\degree$.

\subsection{Interacting potentials}
\label{sect:pot}

The total potential energy
of each system
formed from
the noble gases atoms X,
(X=Ne, Ar, Kr, Xe or Rn),
we have considered
is given by
\begin{equation}
\label{eq:potential}
U_{\rm X} = \sum_{i<j}^{N_{\rm X}} V_{\rm X}(r_{ij}) + \sum_i^{N_{\rm X}}
\sum_j^{N_{\rm C}} V_{\rm X-C}
(r_{ij}) + \sum_{i,j,k}^{N_{\rm C}} V_{ijk},
\end{equation}
where
$N_{\rm X}$ is the number of atoms of a given noble gas X,
$N_{\rm C}$ is the number of carbon atoms in the substrate,
$r_{ij} = |\vec{r_i} - \vec{r_j}|$ and $\vec{r_i}$ is the
position of
the $i$-th atom.
The first term on the right-hand side of \eq{eq:potential} describes the
pair interaction of atoms X of the adlayer, the second, the interaction
of X with the substrate.
The last term models interactions between the carbon
atoms of the substrate.
$V_{ijk}$ is related to the Tersoff potential
\cite{ter88,ter89,ter90} which takes into account not only two-body
interactions but also those of three-body among the carbon atoms in the
substrate.

\begin{figure}[!htb]
\centering
\includegraphics[angle=-90,width=0.8\linewidth]{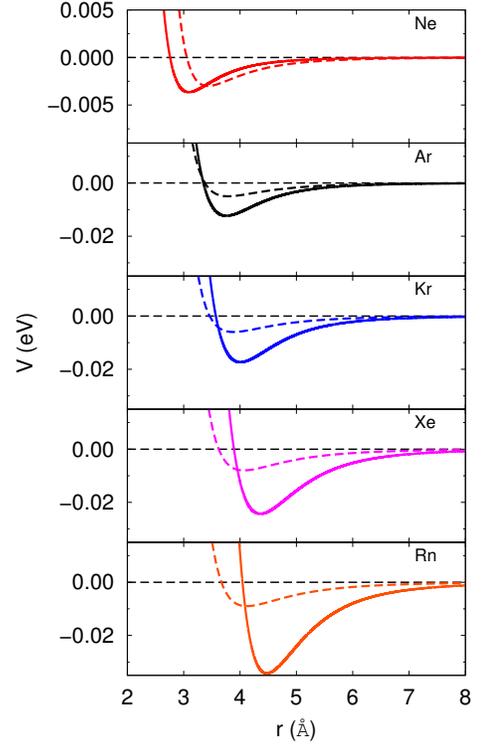}
\caption{\label{fig:pot}(Color online)
$V_{\rm X}$ (solid curves) and $V_{\rm X-C}$ (dashed curves) pair interaction energies.
The Ne-Ne, Kr-Kr
and Xe-Xe curves are for the interatomic potential HFD-B
\cite{azi89,azi86,azi86_2}; the Ar-Ar interactions is given by the HFDID1
potential \cite{azi93}; the Rn-Rn interactions were based on the
Tang-Toennies potential \cite{tan03}.
$V_{\rm X-C}$ is a LJ potential for the
 Ne-C, Ar-C, Kr-C, Xe-C and Rn-C interactions
with the parameters of \tab{xc}.}
\end{figure}

The pair interactions
$V_{\rm X}$
we use
between the adatoms
give an accurate description of their potential energies.
The potential depth and the position
of the minimum increase with increasing atomic number, as illustrated in
Fig. \ref{fig:pot}.

\begin{table*}[!htb]
\centering
\caption{
Parameters
of the Lennard-Jones interaction
for the given X atoms
displayed in the lines of the
table,
$r$ values
where
$V_{\rm X}(r)$ and $V_{\rm X-C}(r)$ are minima
are denoted by
$r_{\rm X}^{(\text{min})}$ and $r_{\rm XC}^{(\text{min})}$,
respectively.}
\begin{tabular}{|l||c|c|c|c|c|c|}
\hline
 & \multicolumn{1}{l|}{$\sigma_{\rm X}$ (\AA)} &
\multicolumn{1}{c|}{$\epsilon_{\rm X}$
(eV)}& $r_{\rm X}^{(\text{min})}$ (\AA)&\multicolumn{1}{l|}{$\sigma_{\rm XC}$ (\AA)} &
\multicolumn{1}{c|}{$\epsilon_{\rm XC}$ (eV)} &
$r_{\rm XC}^{(\text{min})}$ (\AA) \\ \hline \hline
Ne \cite{azi89}   & 2.759 & 0.0036 & 3.091 & 3.055 & 0.003 & 3.429 \\
\hline
Ar \cite{azi93}   & 3.400 & 0.0103 & 3.757 & 3.375 & 0.005 & 3.788 \\
\hline
Kr \cite{azi86}   & 3.571 & 0.0173 & 4.008 & 3.460 & 0.006 & 3.884 \\
\hline
Xe \cite{azi86_2} & 3.892 & 0.0243 & 4.363 & 3.621 & 0.008 & 4.064 \\
\hline
Rn \cite{tan03}   & 3.988 & 0.0343 & 4.477 & 3.669 & 0.009 & 4.118 \\
\hline
\end{tabular}
\label{tab:xc}
\end{table*}

The
pairwise interaction
we employ
between an adsorbate atom X
and the substrate carbon atoms,
$V_{\rm X-C}$ in \eq{eq:potential},
is modeled by a
Lennard-Jones (LJ) potential
\begin{equation}
V_{\rm X-C}(r_{ij}) = 4\epsilon_{\rm XC}
\left[
\left( \frac{\sigma_{\rm XC}}{r_{ij}} \right)^{12}
- \left( \frac{\sigma_{\rm XC}}{r_{ij}} \right)^6 \right],
\end{equation}
where
the parameters of the mixed
pair X-C
is obtained by
combining rules \cite{bru07} of the
X and C atom pairs
\begin{eqnarray}
\label{eq:combining}
\sigma_{\rm XC} &=& \frac{(\sigma_{\rm X}+\sigma_{\rm C})}{2}, \nonumber \\
\epsilon_{\rm XC} &=& \sqrt{\epsilon_{\rm X} \ \epsilon_{\rm C}}.
\end{eqnarray}
The chosen carbon-carbon LJ parameters \cite{inu08} are $\sigma_{C}=3.35$
\AA \ and $\epsilon_{C}=0.0024$ eV. Table \ref{tab:xc} summarizes
$\sigma_{\rm X}$ and $\epsilon_{\rm X}$ from the literature, as well as the
obtained X-C parameters. The corresponding LJ potential energies for the
X-C pairs are
plotted in Fig. \ref{fig:pot}. We note that the interactions between
adatoms and the substrate (X-C) are much weaker than the one for
the noble gases atoms (X-X), except for Ne.

\subsection{Simulation setup}
\label{sect:sim}

Properties of the noble gases adsorbed on graphene systems were determined
using molecular dynamics (MD) in the canonical ($NVT$) \textit{ensemble}.
This classical approach is well-suited for heavy noble gases, with the
possible exception of Ne, where quantum effects might affect some
of the properties in which we are interested. Nevertheless we will disregard
any quantum effect the Ne system might present.

Thermal averages of physical quantities were formed from as many as $10^7$ time 
steps of 1 fs. In the presented study the LAMMPS (Large-scale Atomic/Molecular 
Massively Parallel Simulator) \cite{LAMMPS} code was used.

\begin{figure}[!htb]
\centering
\includegraphics[width=0.5\linewidth]{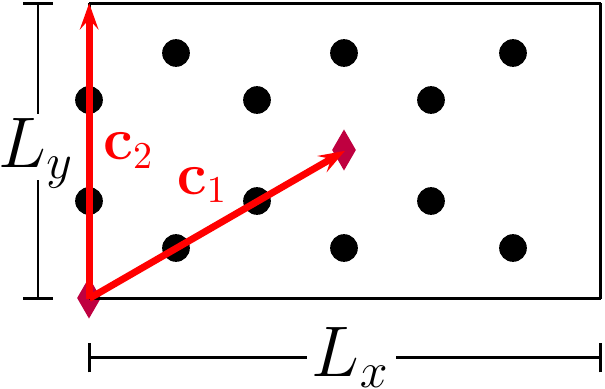} 
\caption{\label{fig:com}(Color online) Commensurate lattice 
$\sqrt{3}\times\sqrt{3}R30\degree$ unit cell. The circles
($\bullet$) represent the position of C atoms, diamonds
($\blacklozenge$)
indicate the position of the heavy noble gases atoms, we
also plot the primitive vectors of \eq{eq:sqrt3}.}  
\end{figure}

The initial positions of the atoms must be chosen carefully. The two
dimensional unit cell of the $\sqrt{3}\times\sqrt{3}R30\degree$
commensurate lattice corresponds to 12 carbon atoms and 2 heavy noble
gases atoms, as shown in Fig. \ref{fig:com}. The lengths of the cell are
$L_x$ = $3 \sqrt{3} \ b$ and $L_y$ = $3 \ b$. The surface density $\rho$
will be given hereafter in units of $\rho_0$=0.0636 atoms/\AA$^2$, the
density of the $\sqrt{3} \times \sqrt{3}$ commensurate lattice. The
initial $z$ coordinate of the noble gases atoms is set as the minimum of
the X-C pair potential.

Initial positions using the
$\sqrt{7}\times\sqrt{7}$ superlattice can be achieved with a cell of 224
X
atoms and 784 C atoms, with corresponding superficial density of
$\rho=1.71$.
Densities other than those of the two commensurate structures were
initialized with triangular lattices with suitable lattice parameters.
Tab. \ref{tab:densities} summarizes the numbers of noble gases and
carbon atoms and corresponding densities employed in the simulations. We
use periodic boundary conditions
(PBC) in the $xy$ plane of the substrate and the $z$ coordinates of the
carbon atoms were kept fixed. No PBC were employed in the $z$ direction,
normal to the substrate, in order to allow evaporation of noble gases
atoms. Because the adlayer may be incommensurate with the substrate
symmetry, we chose certain surface densities which guarantee the adatoms
in a triangular lattice. The initial velocities were chosen from a
Gaussian distribution compatible with the desired temperature.

\begin{table}%[!htbp]
\centering
\caption{Densities, number of noble gases atoms $N_{\rm X}$, and
carbon atoms $N_{\rm C}$  used
in the simulations.}
\begin{tabular}{|r||l|l|}
\hline
\multicolumn{1}{|c||}{$\rho$} & $N_{\rm X}$ & $N_{\rm C}$ \\ \hline \hline
0.38 & 210 & 3360 \\ \hline
0.73 & 232 & 1920 \\ \hline
1.00 & 224 & 1344 \\ \hline
1.08 & 216 & 1200 \\ \hline \hline
1.71 & 224 & \phantom{0}784 \\ \hline
\end{tabular}
\label{tab:densities}
\end{table}

\section{Results}
\label{sect:res}

Our results are mainly for
Ar/graphene systems, which was motivated in part by the
interesting behavior of the relate system Ar/graphite. We studied
properties at four Ar surface densities; at 
$\rho=1$ that in our units corresponds to an adlayer in the
$\sqrt{3}\times \sqrt{3}$
commensurate lattice,
one slightly above this value, at $\rho=1.08$, and two
below: $\rho=0.73$ and 0.38. Since the system Ne/graphite exhibits
the superlattice
$\sqrt{7}\times\sqrt{7}$,
we also considered for our Ne/graphene system the corresponding density 
\textsl{i.e.}, 
$\rho=1.71$, in addition to $\rho=1$.
% for the superlattice $\sqrt{7}\times\sqrt{7}$.
For Xe, Kr and Rn we restricted ourselves to the
% commensurate lattice $\sqrt{3}\times \sqrt{3}$,
density  $\rho=1$.

\subsection{Commensurate structures}
\label{sect:com_res} 

\begin{figure}[!htb]
\centering
\subfigure[]{
  \includegraphics[angle=-90,width=0.475\linewidth]{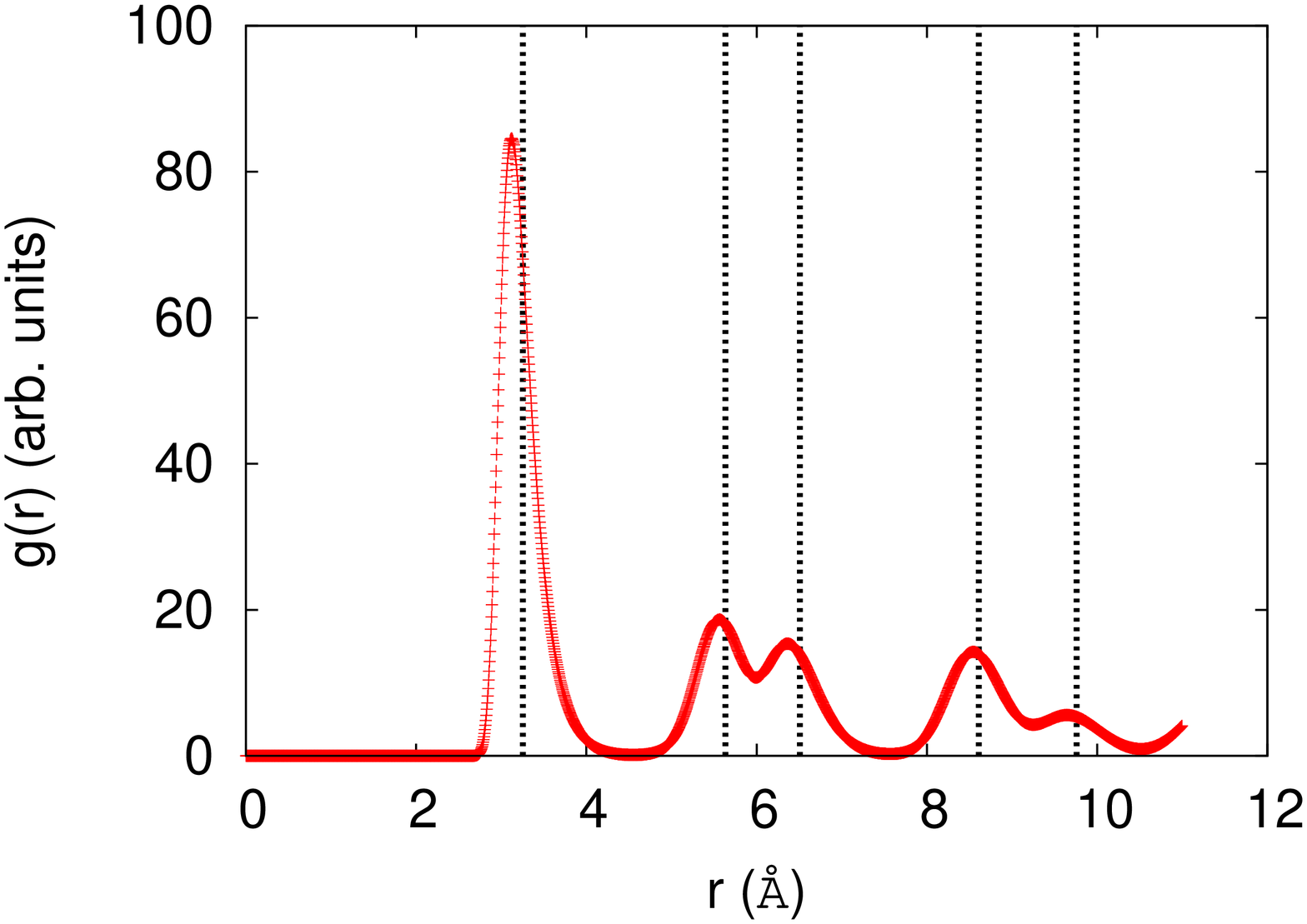}}
%\hspace{.02\linewidth}
\subfigure[]{
  \includegraphics[angle=-90,width=0.475\linewidth]{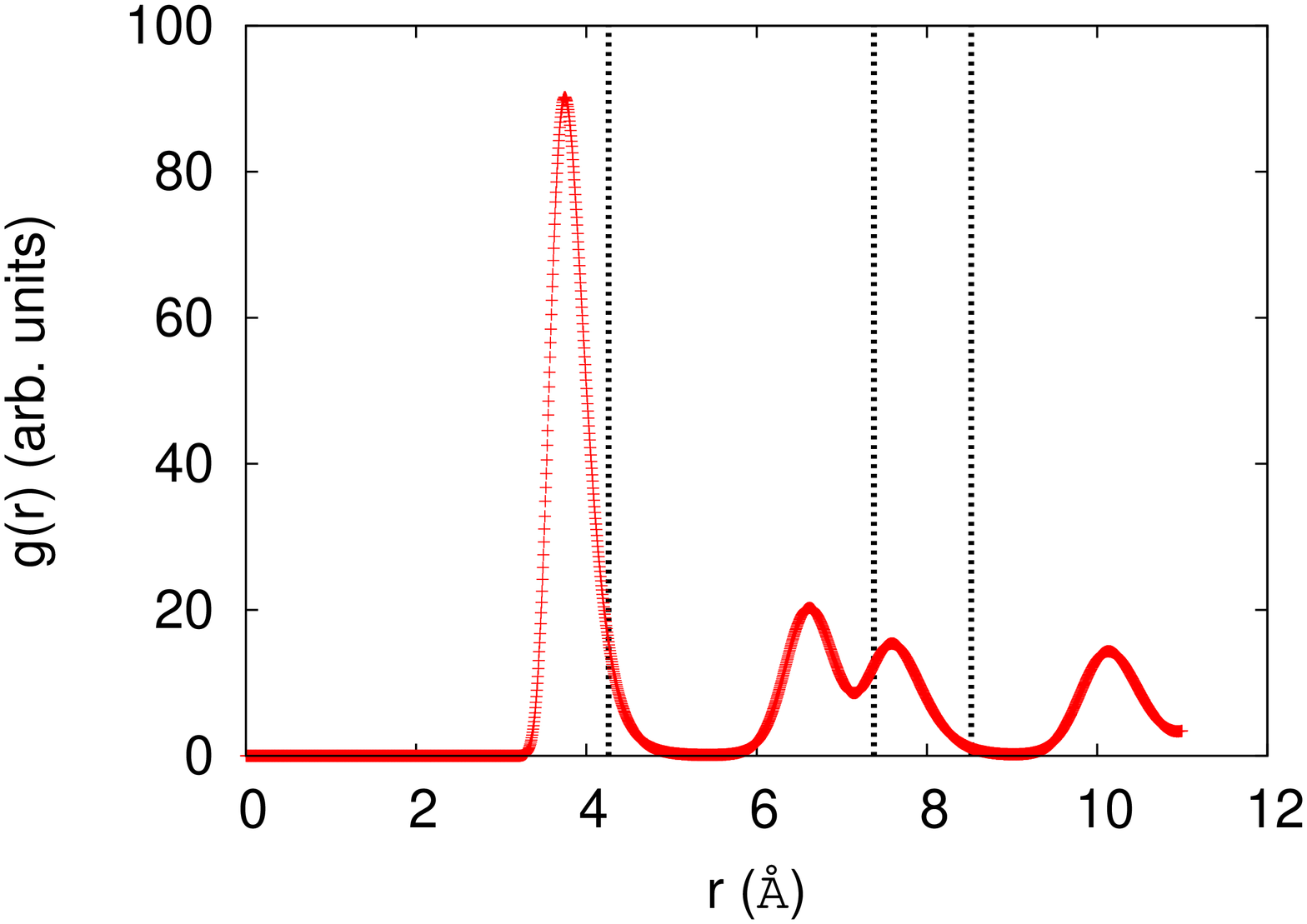}}
\caption{\label{fig:gr_ne}(Color online)  Radial pair distribution functions 
for Ne atoms at $T=$ 10 K and $\rho=1.71$
(a),
and for the
Ar adlayer at
$\rho=1$ and T=40 K
(b).
The dashed lines
correspond to the neighbors distances of atoms in the ideal sites of the
commensurate lattices, $\sqrt{7} \times \sqrt{7}$ (a) and 
$\sqrt{3} \times \sqrt{3}$ (b).}
\end{figure}

One of the characteristics of adsorption on solid surfaces is the 
possibility of formation of commensurate layers.
The radial pair distribution function $g(r)$ is a useful tool to investigate 
this property.
The atomic
distances
of a two dimensional
ideal commensurate structure
are easily
calculated. We computed $g(r)$ of the heavy noble gases on
the graphene substrate and compared our results with the ideal structures.
Typical examples of our results are shown in
\fig{gr_ne}.
For the commensurate layers,
the radial pair distribution function presents peaks in good
agreement with a
commensurate lattice for a wide range of temperatures that go
almost to the melting temperature.
The Ne adlayer shows this behavior,
in
Fig.~\ref{fig:gr_ne} (a)
at $\rho=1.71$ and $T=10$ K,
a temperature well below melting $T_m=25.5$ K,
we can see that the system is in a
$\sqrt{7}\times\sqrt{7}$ superlattice.
In the following sections we will discuss in detail how the melting
temperatures were computed.
As the temperature
increases towards the melting temperature, in general the peaks broaden.
On the other hand, for non-commensurate lattices of the adlayer with respect to the
graphene substrate,
the atoms are not found preferentially at
distances compatible with the commensurate structure
for any considered temperature or density, as we show in
Fig.~\ref{fig:gr_ne} (b) for the Ar adlayer.

\begin{figure}[!htb]
\centering
\includegraphics[angle=-90, width=0.8\linewidth]{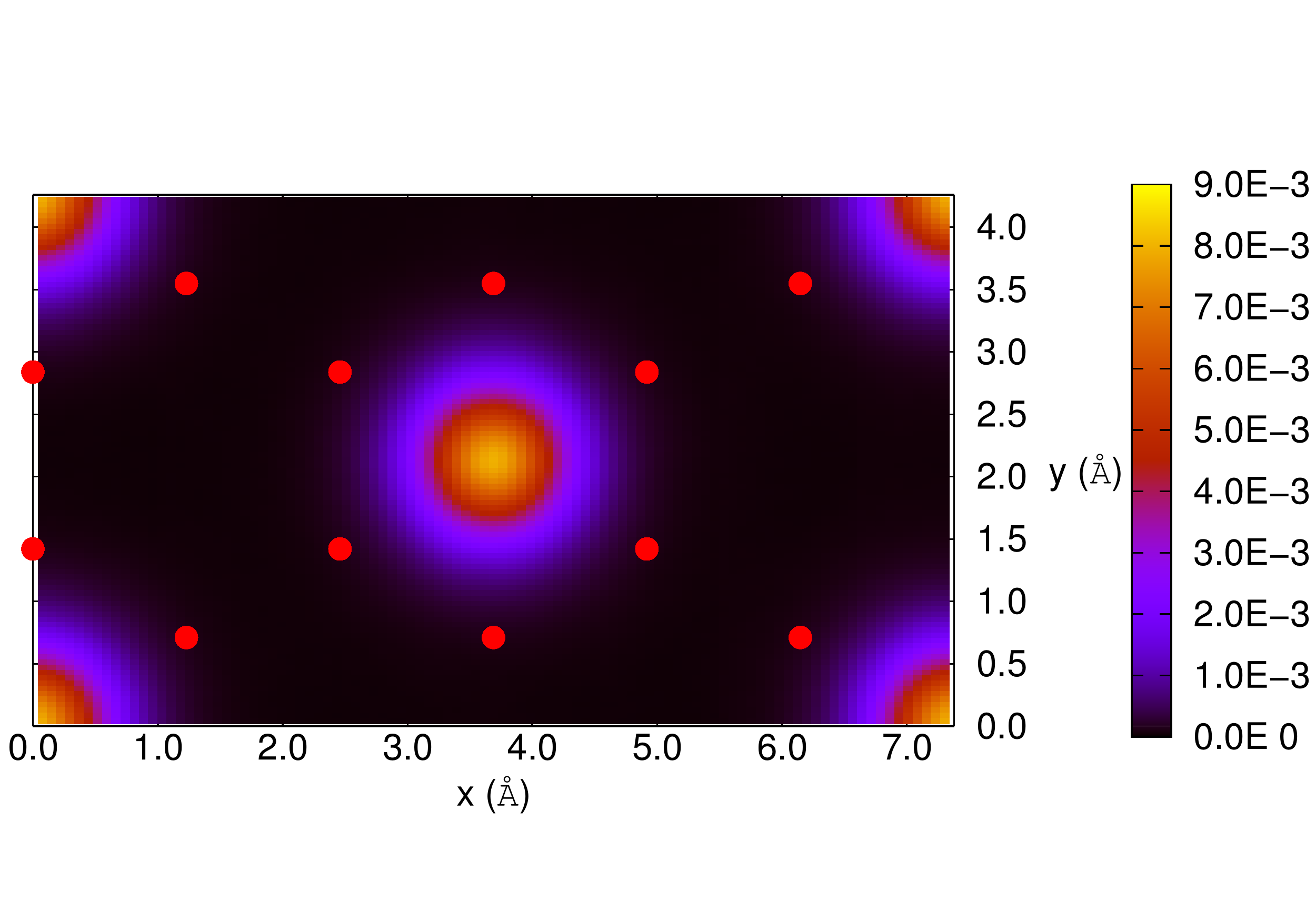}
\caption{\label{fig:prob_kr}(Color online)  Spatial distribution of Kr
atoms over a $\sqrt{3} \times \sqrt{3}$ cell below melting.
The red dots represent the ideal positions of the
carbon atoms.}
\end{figure}

We also analyzed the spatial distribution of rare-gas atoms over the
graphene sheet by accumulation of a two-dimensional histogram in a grid of
the $xy$ plane over the unit cells of the system depicted in
Fig.~\ref{fig:com}.
The results for the two-dimensional histogram of krypton atoms at $\rho=1$ and
$T=$ 100~K are presented in
\fig{prob_kr}.
As we see, the most probable locations of
the Kr atoms are the lattice sites of the
$\sqrt{3} \times \sqrt{3}$ structure.
Further evidence to
support our claim
that Kr atoms form
a commensurate
adlayer comes from
a plot of $g(r)$. It shows
peaks in agreement with
nearest neighbors distances of the ideal
commensurate structure in a fashion
similar to what is depicted in \fig{gr_ne} (a). This behavior is
observed for temperatures below melting, $107.9$~K. At higher temperatures,
above melting, the
probability of finding a Kr atom over the center of a substrate hexagon is
higher, although it can be found in other positions as well. 

By calculating the radial pair distribution function
of atoms in adlayers of Ar, Xe and Rn
we found no evidence of
commensurate structures with the substrate,
even at low temperatures.
Nevertheless,
the probability of
finding one adatom over the center of a substrate hexagon is still larger than
in other locations.
On top of the carbon atoms
were the locations where
atoms of the noble gases were found with lowest probability.

\subsection{Adlayer Specific Heat}
\label{sect:cv}
	
The specific heat $c_v$ is a thermodynamical quantity of both experimental
and theoretical interest. It is computed by considering the fluctuations
of the average internal energy
$\langle \delta E^2 \rangle_{\rm NVT}$
of the adatoms

\begin{equation}
\label{eq:cv}
c_v = \frac{\langle \delta E^2 \rangle_{\rm NVT}}{N_{\rm X} k_B T^2},
\end{equation}	
in the canonical ensemble.

\begin{figure}[!htb]
\centering
\includegraphics[angle=-90,width=0.8\linewidth]{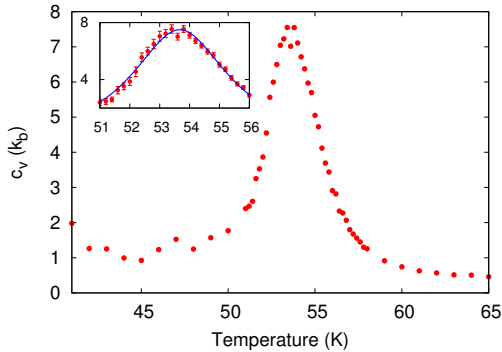}
\caption{\label{fig:cv108}(Color online)  The specific heat of Ar
as a function of the temperature for $\rho=1.08$. The region of the peak
is magnified in the inset
together with the errorbars,
the solid line is a function of the Lorentzian form fitted to the
results}
\end{figure}

The specific heat as a function
of the temperature for Ar at $\rho=1.08$
that we have calculated is shown in
Fig.~\ref{fig:cv108}.
The behavior of the
specific heat for the other noble gases and densities is qualitatively the
same.
The peak temperature and its characteristic width were
determined by fitting to the computed points a curve of the Lorentzian form,
\begin{equation}
\label{eq:cv_lor}
c_v(T) = \frac{a}{\pi \gamma \left[ 1 + \left( \frac{T-T_0}{\gamma} \right)^2
 \right]} + b,
\end{equation}
where $T_0$ is the location of the peak, $2\gamma$ is its FWHM, $a$ is
related to the area of the curve and $b$ is a $y-$shift. We carried out
this procedure to the other rare-gases and densities; the results are
summarized in Table~\ref{tab:sh}.

\begin{table}[!htb]
\centering
\caption{Temperature and width of the specific heat peak for the given noble 
gases and densities.}
\begin{tabular}{|c|c|c|c|}
\hline
 & \multicolumn{1}{c|}{$\rho$} & $T_0$ (K) & FWHM (K) \\ \hline \hline
\textbf{Ne} 	& 1.00 & \phantom{1}15.7 $\pm$ 0.1 & \phantom{1}1.9 $\pm$\phantom{1}0.2 \\ \hline
 			& 1.71 & \phantom{1}35.6 $\pm$ 0.2 & \phantom{1}8.3 $\pm$\phantom{1}0.8 \\ \hline \hline
\textbf{Ar} 	& 0.38 & \phantom{1}53.6 $\pm$ 0.2 & \phantom{1}8.6 $\pm$\phantom{1}0.7 \\ \hline
 			& 0.73 & \phantom{1}54.4 $\pm$ 0.1 & \phantom{1}3.7 $\pm$\phantom{1}0.3 \\ \hline
 			& 1.00 & \phantom{1}55.1 $\pm$ 0.1 & \phantom{1}3.9 $\pm$\phantom{1}0.3 \\ \hline
 			& 1.08 & \phantom{1}53.7 $\pm$ 0.1 & \phantom{1}4.0 $\pm$\phantom{1}0.2 \\ \hline \hline
\textbf{Kr} 	& 1.00 & 107.4 $\pm$ 0.2 & \phantom{1}6.5 $\pm$\phantom{1}1.2 \\ \hline \hline
\textbf{Xe} 	& 1.00 & 134.3 $\pm$ 0.5 & \phantom{1}7.9 $\pm$\phantom{1}1.3 \\ \hline \hline
\textbf{Rn} 	& 1.00 & 188.9 $\pm$ 1.6 & 41.0 $\pm$ 8.9 \\ \hline 
\end{tabular}
\label{tab:sh}
\end{table}

The temperatures of the specific heat peaks
increase with the atomic number
as we can see
at the density $\rho=1$.
Additionally, we observe an enlargement of their
characteristic widths. These properties are consistent with
the bulk properties of these systems.

It is interesting to compare the specific heat peaks of Ar adlayers at different
densities. The temperature $T_0$ of the peak
has its
maximum value at the density $\rho=1$. This behavior
might be credited to the effects of the graphene
substrate that are more evenly distributed through the adatoms. 
Although the Ar
atoms of the
adlayer are not preferentially found in the ideal commensurate lattice,
the substrate favors the expansion of the adlayer
towards the $\sqrt{3}\times \sqrt{3}$ structure.
If we consider the statistic
uncertainty of the specific heat peak, we possibly could include in our
analysis the system at the density
$\rho=0.38$.
However, because of size effects,
we chose not to do this.
At this low density we have found that the adlayer forms islands where
the proportion of atoms in the border
to those inside are substantial. This is a situation that
might cause deviations in the observed trend as we can see in
the obtained FWHM.
	
\subsection{Melting}
\label{sect:melt} 		
	
The melting of the adlayers is related to the loss of sixfold
symmetry that is present in the solid phases of these systems. 
The study of their melting can be done by following the evolution
with temperature
of an
order parameter related to the sixfold symmetry
For this purpose
we introduce the order parameter
\begin{equation}
\label{eq:psi}
\Psi_6 = \frac{1}{N_B} \left< \Bigg| \sum_{j,k_j}^N
\exp{(6i\Phi_{jk_j})} \Bigg| \right> ,
\end{equation}
where $\Phi_{jk_j}$ is the angle between the projections in the $xy$ plane
of the relative position of atom \textit{j} and its nearest neighbors $k_j$ 
with respect
to a fixed axis, \textsl{e.g.}, the $x$-axis
in this plane, and $N_B$ is the number of bonds used in the
calculation. The sum on \textit{j} extends over all noble gas atoms.
The brackets indicate a thermal average. If the adlayer possesses six-fold
symmetry, as a triangular lattice does, $\Psi_6 = 1$. For an isotropic
fluid,
$\Psi_6$ is zero.
Moreover, the loss of the sixfold symmetry can be
characterized by a peak in the susceptibility $\chi_6$ of $\Psi_6$ given
by
\begin{equation}
\label{eq:chi}
\chi_6 = \frac{ \langle \Psi_6^2 \rangle - \langle \Psi_6 \rangle^2 }{T}.
\end{equation}

\begin{figure}[!htb]
\centering
\includegraphics[angle=-90,width=0.8\linewidth]{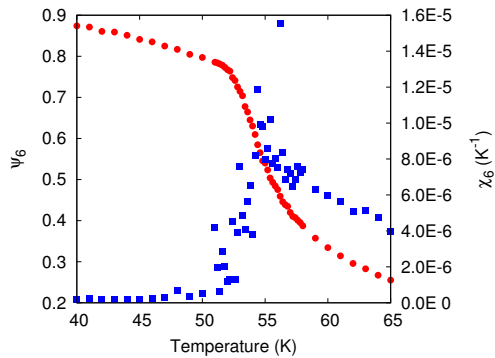}
\caption{\label{fig:order}(Color online)  Order parameter (circles) and 
susceptibility
(squares) for Ar at $\rho=1.08$ as a function of the temperature.}
\end{figure}

The order parameter $\Psi_6$ and its susceptibility $\chi_6$ for Ar at
$\rho = 1.08$ is shown in
Fig.~\ref{fig:order}.
The order parameter is near unity for low
temperatures and drops  sharply
when the melting occurs. The susceptibility
peaks near the transition temperature.
We have observed a similar behavior
at other densities for this system and for the others rare-gases adlayers
as well.

\begin{table}[!htb]
\centering 
\caption{Melting temperature and characteristic width for various noble gases and densities.}
\begin{tabular}{|c|c|c|c|}
\hline
 & \multicolumn{1}{c|}{$\rho$} & $T_m$ (K) & FWHM (K) \\ \hline \hline
\textbf{Ne}	& 1.00 & \phantom{1}15.9 $\pm$ 0.1 	& \phantom{1}1.3 $\pm$ 0.2 \\ \hline
 			& 1.71 & \phantom{1}25.5 $\pm$ 0.2 	& \phantom{1}9.5 $\pm$ 1.5 \\ \hline \hline
\textbf{Ar} 	& 0.38 & \phantom{1}53.9 $\pm$ 0.1 	& \phantom{1}8.2 $\pm$ 1.2 \\ \hline
 			& 0.73 & \phantom{1}55.2 $\pm$ 0.1 	& \phantom{1}4.0 $\pm$ 0.8 \\ \hline
 			& 1.00 & \phantom{1}56.0 $\pm$ 0.1 	& \phantom{1}3.1 $\pm$ 0.5 \\ \hline
 			& 1.08 & \phantom{1}55.5 $\pm$ 0.2 	& \phantom{1}5.2 $\pm$ 2.0 \\ \hline \hline
\textbf{Kr} 	& 1.00 & 107.9 $\pm$ 0.2 			& \phantom{1}4.2 $\pm$ 0.6 \\ \hline \hline
\textbf{Xe} 	& 1.00 & 175.3 $\pm$ 0.4 			& 10.0 $\pm$ 1.5 \\ \hline \hline 
\textbf{Rn} 	& 1.00 & 234.9 $\pm$ 1.0 			& 20.0 $\pm$ 5.2 \\ \hline
\end{tabular}
\label{tab:melt}
\end{table}	

The points of the susceptibility $\chi_6$ were fitted
to a function of the Lorentzian form. We interpret the fitted parameter
$T_0$
of \eq{eq:cv_lor},
as the melting temperature $T_m$.
Our results are presented in \tab{melt}.
If we compare different systems at the same density $\rho=1$, the melting
temperature and the Lorentzian FWHM
increase with the atomic number.
Specifically for the Ar adlayers we can analyze the melting temperature
as a function of the density. This quantity
has a maximum at $\rho=1$,
and the behavior as a function of the density
might be attributed to
the substrate.
Like what we 
observed for the
specific heat,
the substrate drives
in a more effective way 
an evenly distribution of the adlayer atoms
towards a
$\sqrt 3 \times \sqrt 3 $ structure.	
	
\subsection{Nearest neighbor distance}	
\label{sect:first}	
	
An estimate of the nearest neighbor distance $\langle a \rangle $ can be
obtained from the pair distribution function $g(r)$. The behavior of
$\langle a \rangle$ as a function of the temperature can be quite
different for the various densities and rare-gases studied.  In fact, we
have observed three different ones.
It can smoothly increase with
temperature, which is the case of Ne at $\rho=1$ and of Ar at all
considered densities. For the two commensurate layers that we have observed
in our calculations,
$\sqrt{7}\times\sqrt{7}$ Ne and $\sqrt{3}\times\sqrt{3}$ Kr, the
nearest-neighbor (NN) distance
presents a minimum near
the melting temperature.  Finally, the NN distance
$\langle a \rangle $
can sharply increase as the
temperature approaches the specific heat peak
in a way that can be associated with the promotion of adatoms to a second
layer, as we have observed for Xe
and Rn at $\rho=1$.

\begin{figure}[!htb]
\centering
\subfigure[]{
\includegraphics[angle=-90,width=0.48\linewidth]{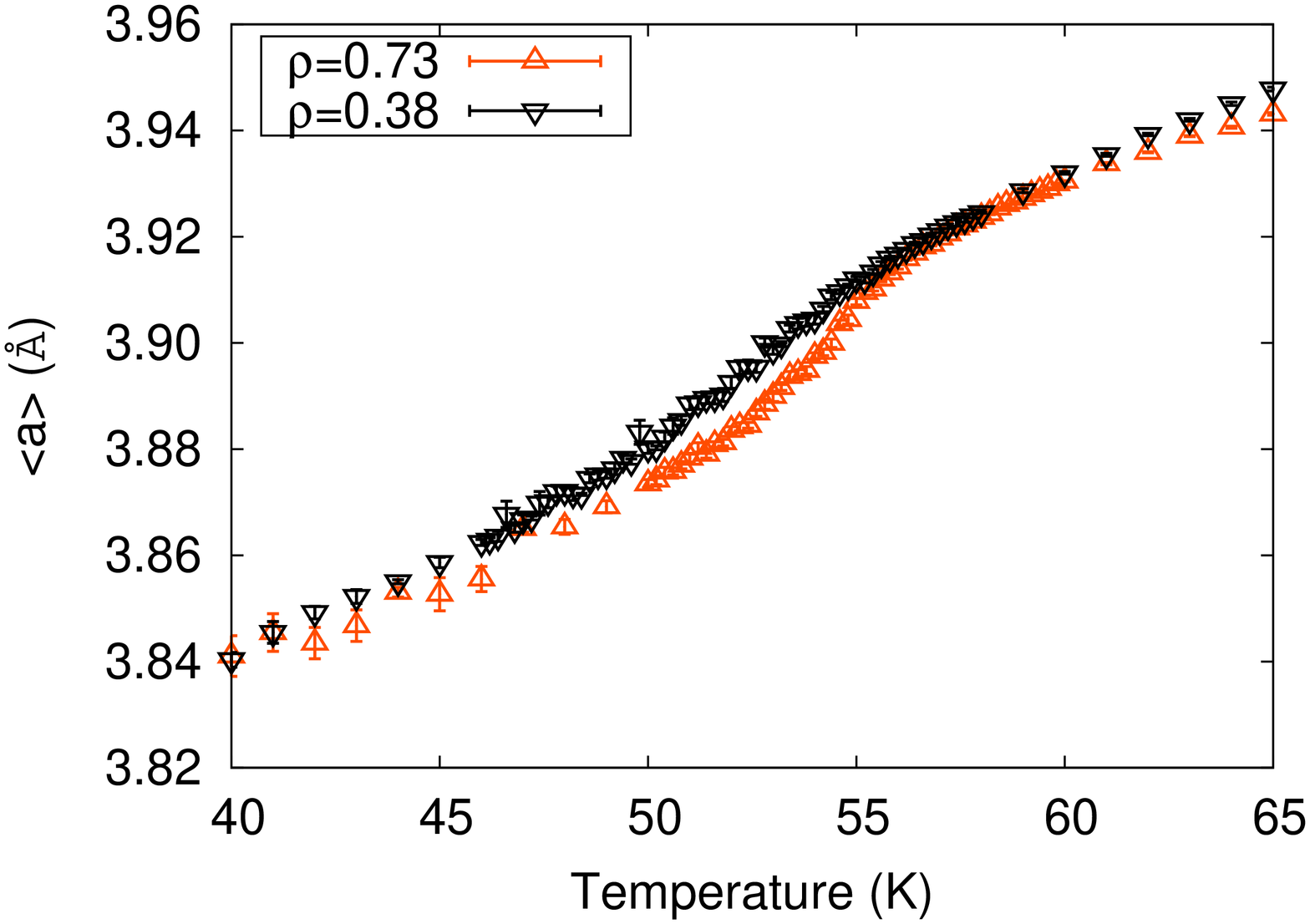}}
\subfigure[]{
\includegraphics[angle=-90,width=0.48\linewidth]{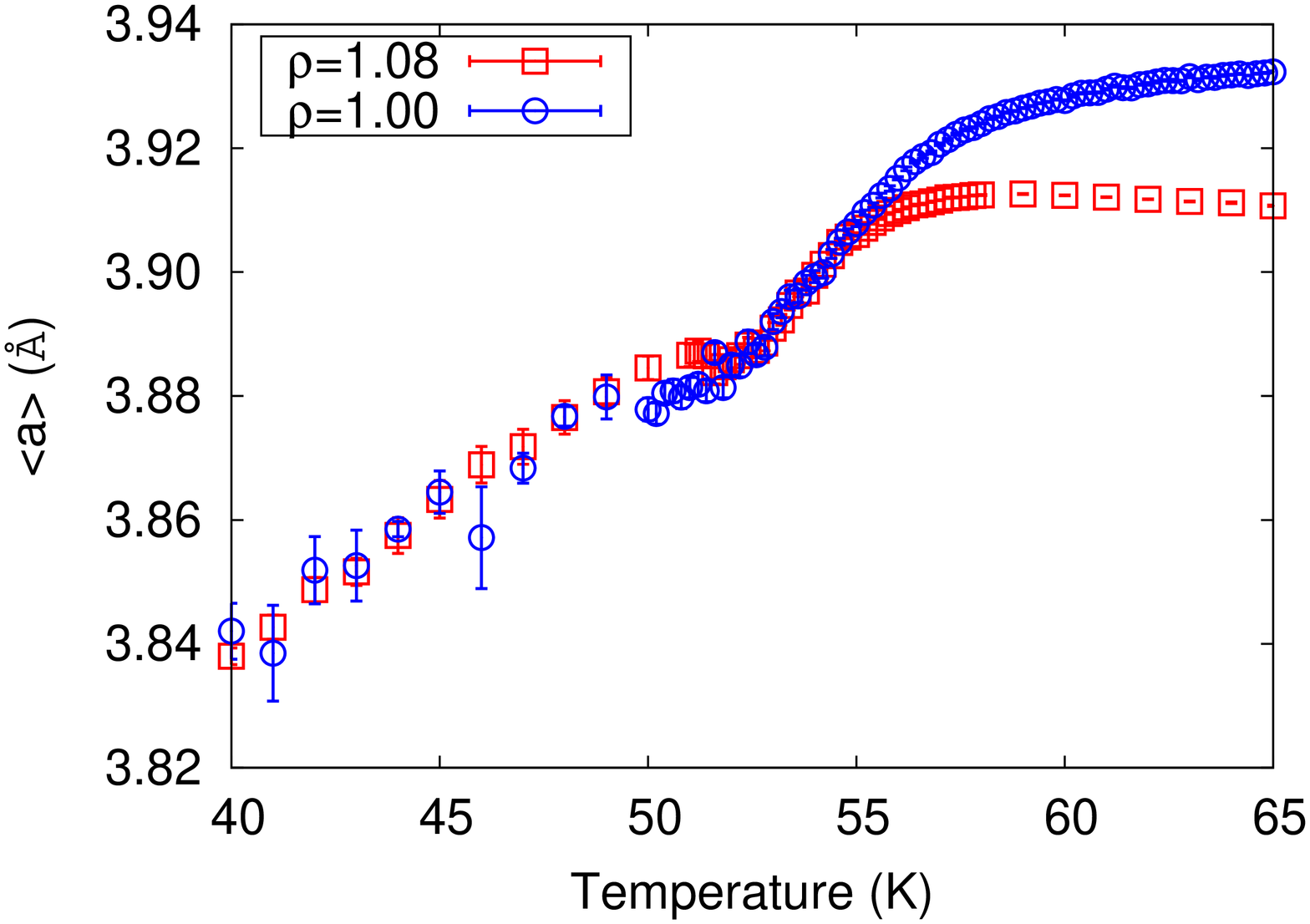}}
\caption{\label{fig:lattice_ar}(Color online)  Nearest neighbor distance as a 
function of the temperature for Ar at the given densities.}
\end{figure}

We begin by analyzing the mean NN distance of Ar adatoms at various
densities as a function of the temperature presented in \fig{lattice_ar}.
At low densities, \textsl{i.e.}, $\rho=0.38$ and 0.73, we observe a NN
distance that increases with temperature. This behavior can be
explained by the fact that, at low densities, the adlayer does not cover
entirely the graphene substrate. However, with the increase of thermal
energy, the atoms become
more free to move, and will occupy more evenly the available space.
At density $\rho=1.08$ and
for the high range of temperatures we have considered in our calculations,
we see that the  NN distance saturates.
At the density $\rho=1$ it seems that the mean value of
the NN distance
is almost reaching a saturation regime
at this same range of temperatures.
This situation can be explained
if we consider the Helmholtz free energy $F$ (we are working within the
canonical \textit{ensemble}),
\begin{equation}
F = E - TS,
\end{equation}
where $E$ is the internal energy and $S$ is the entropy. For low
temperatures, the free energy is dominated by the internal energy $E$, and
the layer can expand as the temperature rises. Since the term $TS$ is
significant at higher temperatures, disorder in the layer is energetically
favored instead of its expansion. This situation of nearly constant
$\langle a \rangle $ would persist until second layer promotion would be
thermally activated \cite{fle06}.
At the two highest
densities of
the Ar adlayers we have
observed a puzzling feature,
for a small range of temperatures
below the specific heat peak,
which is the approximated constant value of $\langle a \rangle $.
This behavior
can not be associated
with the promotion of atoms to a
second layer as we will see in Sec. \ref{sect:z}. 
%by
%studying the average distance of
%the atoms in the adlayer to the
%substrate.
%
For Ne at $\rho=1$ the NN distance
increases with the temperature, qualitatively displaying a behavior that 
resembles the one
of Ar at $\rho=0.78$.

\begin{figure}[!htb]
  \centering
  \includegraphics[angle=-90,width=0.7\linewidth]{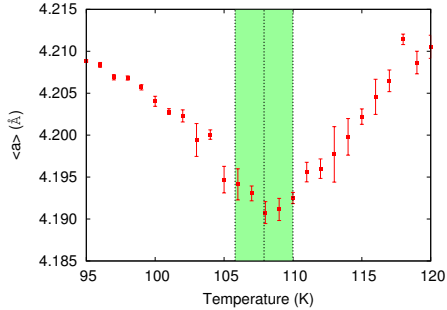}
  \caption{(Color online) Nearest neighbor distance of Kr atoms in adlayers at 
  $\rho=$ 1.
The centered dotted vertical line shows
the melting temperature $T_0=107.9$ K.
The region between the external dotted vertical
lines correspond to the FWHM of the
melting peak.}
  \label{fig:lattice_kr}
\end{figure}

The second behavior we have observed is typical of adlayers commensurate
with a substrate,
where the nearest neighbor distance
initially decreases with temperature before increasing as expected.
In our calculations,
Ne at
$\rho=1.71$ and Kr at $\rho=1$ have presented
this behavior.
For Kr we
show in
Fig. \ref{fig:lattice_kr}
a plot of the average
NN distance as a function of temperature.
The
minimum of the nearest neighbor distance
happens about the
melting temperature,
and the FWHM of the fitted curve to $\chi_6$
encompasses the change between negative and positive slopes.
Above $T_0$ the liquid expands, as expected. 
For the Ne adatoms,
as the temperature increases they gain
sufficient energy to overcome the substrate potential and get closer
together. On the other hand, the Kr adatoms feel more the substrate, as the
temperature increases up to melting, and the NN distance decreases.

\begin{figure} %[!htb]
\centering
\includegraphics[angle=-90,width=0.7\linewidth]{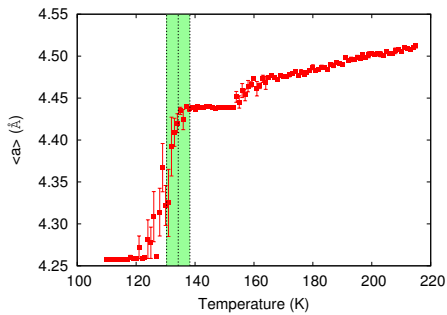}
\caption{\label{fig:lattice_xe}(Color online) First neighbor distance as a
function of the temperature for Xe at $\rho=1$. The dotted vertical lines
show the FWHM temperature region of the specific heat peak indicated by a
dotted vertical line at $T_0 = 134.3$ K}
\end{figure}

Finally, a third behavior is observed for adsorbed layers of Xe
and Rn.
The NN distance as a function of
temperature for
Xe at $\rho=1$
is shown in
Fig.~\ref{fig:lattice_xe}.
The abrupt change of this quantity
happens near the specific heat peak, as we can
see in the figure, and a substantial part of the step variation in the NN
distance occurs within a temperature range given by the FWHM of this peak.
The promotion of Xe atoms to a second layer is responsible for the
observed behavior.
The Rn adlayer at $\rho=1$ presents similar features to those we have
described above.

\subsection{Distance from the substrate}
\label{sect:z}

\begin{figure}[!htb]
\centering
\includegraphics[angle=-90,width=0.7\linewidth]{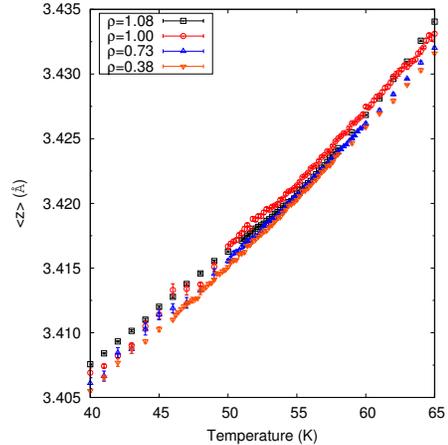}
\caption{\label{fig:z_ar}(Color online) Distance from the substrate as a 
function of the temperature for Ar adlayers at the given densities.}
\end{figure}

The average perpendicular distance
$\langle z \rangle$
from the adatoms to
the substrate plane was computed by accumulating the $z$ coordinates of
the adatoms during the simulations. This quantity for Ar adlayers at
$\rho=0.38$, 0.73, 1 and 1.08 is displayed in Fig. \ref{fig:z_ar}. The
expected behavior of the distance
$\langle z \rangle$
from the substrate,
namely, its increase
with density is
observed at low and high temperatures.
However near the melting temperature and
the specific heat peak deviations occur.
For example, at $\rho=1$,
$\langle z \rangle$
presents its
highest value.
In this work we have no evidence of a second
adlayer formation
of Ar atoms.

\begin{figure}[!htb]
\centering
\includegraphics[angle=-90,width=0.7\linewidth]{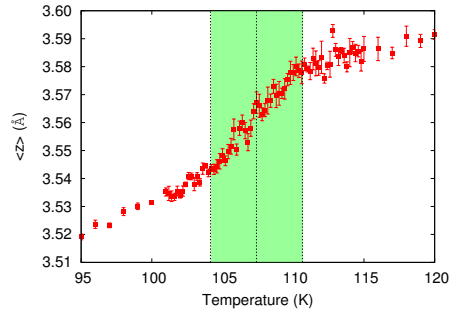}
\caption{\label{fig:z_kr}(Color online) Average distance from the Kr
adlayer to the graphene substrate as a function of the temperature at
density $\rho=1$.  The vertical dotted lines show the specific heat peak
at $T_0=$ 107.4 K and the temperature region of the FWHM of this peak.}
\end{figure}

Adlayers of Ne present similar characteristics to those of Ar. However for
the adlayer at density $\rho=1$ the adatoms are always closer to the
substrate than those at
$\rho=1.71$, most
probably due to a much smaller coverage in this case. On the other hand, the
Kr adlayer presents a fascinating behavior that can be seen in Fig
\ref{fig:z_kr}, the distance from substrate has an inflection point, which
coincides with the specific heat peak temperature.

\begin{figure}[!htb]
  \centering
  \includegraphics[width=0.7\linewidth]{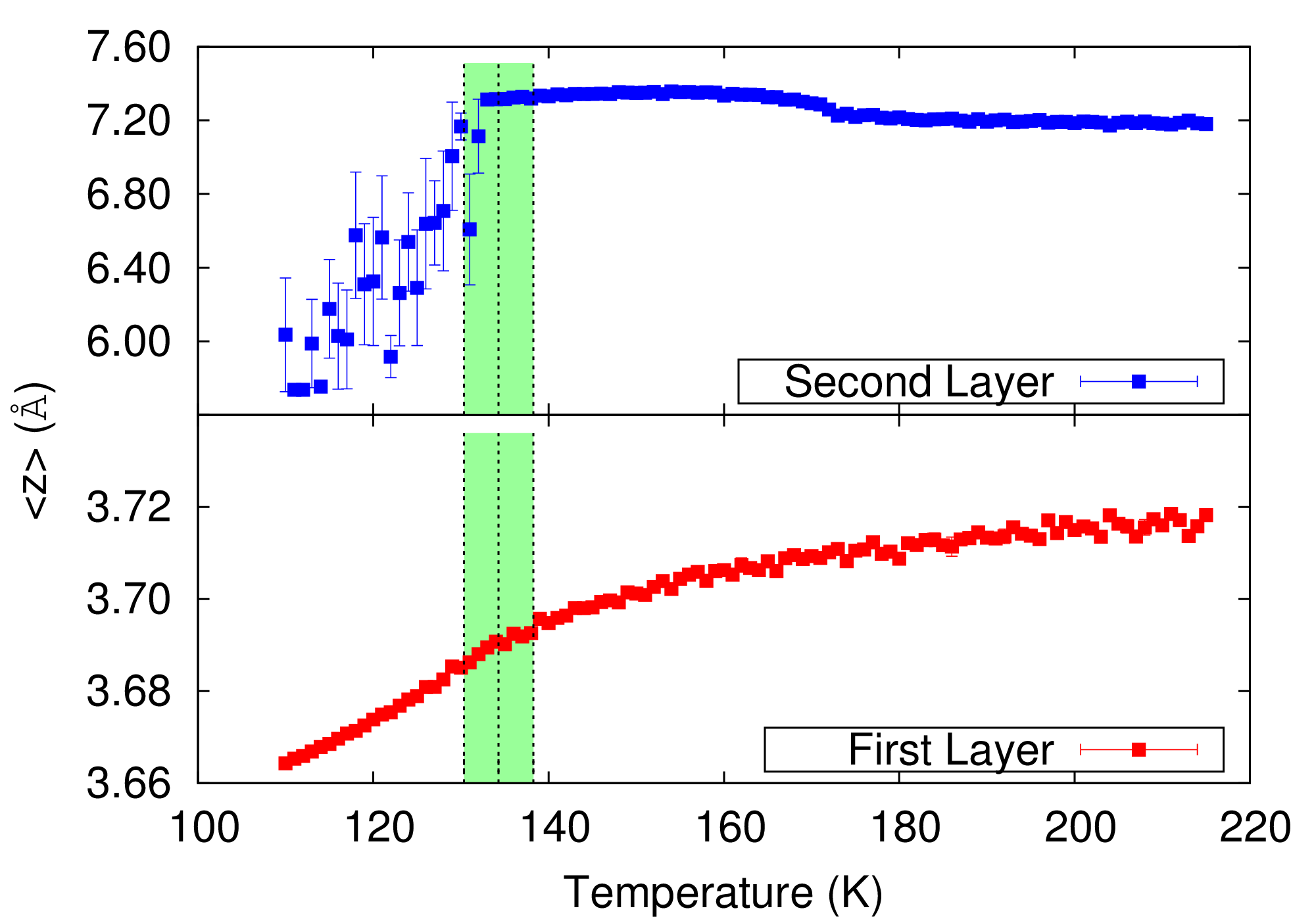}
  \caption{(Color online) Distance from the substrate as a function of the
temperature for Xe adlayers at $\rho=$ 1. The dotted vertical lines
corresponds to the specific heat peak $T_0=134.3$ K and also delimits
its corresponding
FWHM. Note that the $y$-scale differs between the different layers.}
  \label{fig:z_xe}
\end{figure}

Finally we analyze the Xe adlayer, where there was the formation of a second 
adsorbed layer as we can see in
Fig \ref{fig:z_xe}.
In the second adlayer
the distance $\langle z\rangle$
is approximately constant
for temperatures above
$T_0$, the temperature of the
specific heat peak.
It is interesting to note that
in the first adlayer the distance
$\langle z \rangle$
continues to increase smoothly even after the second layer becames stable.

\section{Discussion and Conclusions}
\label{sect:disc} 

One of the contributions of our work on adsorption of noble gases on
graphene is the evidences of commensurate adlayers, which depend strongly
on the symmetry of the exposed substrate surface. Since graphene and
graphite present the same surface, we looked for the well-known
commensurate structures previously observed on graphite \cite{bru97}.
We observed two
commensurate structures at low temperatures; the Ne adlayer forms a
$\sqrt{7}\times \sqrt{7}$ lattice and Kr adatoms are found in the
$\sqrt{3}\times\sqrt{3}$ structure. Although both substrates share many
aspects, this is remarkable because graphene is a much less attractive
substrate than graphite, and we could not know if these structures would
be found \textit{a priori}.
However, we were not able to observe a
commensurate phase for Xe adlayers like it was observed for this system in
a graphite substrate.

The understanding of the specific heat peaks and the melting transition
can be attained by considering the influence of the substrate on the
adlayer.
The specific heat is a quantity of thermodynamical interest that can be
measured in experiments with relatively simple apparatus. Its peaks can
hint the order of the melting transition. For graphite substrates, all
heavy noble gases present continuous melting, with the possible exception
of Ar due to a narrow peak in the specific heat (at the shoulder of a
broad peak at $\sim$ 50 K). We determined the position and width of the
peaks for the heavy rare-gases at various densities, and we only found
evidences of continuous melting, due to the broad FWHM observed for these
peaks.

The melting, and its sixfold symmetry, was also investigated. We
introduced an order parameter and determined the melting temperature using
its susceptibility.  The melting temperature, along with the FWHM,
indicates a temperature range in which we have solid-liquid coexistence.
Moreover, the probability of finding
one noble gas atom over the center of a carbon hexagon is higher than in
other positions. This might explain the deviations found for Ar at
$\rho=1.08$ in the specific heat and also in the melting transition.
Apparently, because of the energy costs, the increase of the density
slightly above $\rho=1$ allows a more frequent occupation
of sites other than the center of the hexagons, which facilitates the
melting transition.

The behaviors of the different adlayer are very rich. In addition to the
arguments we already gave to explain our observations we can advance some
tentative ideas based on the interplay between the mutual interaction of
the adlayer atoms and their interaction with the substrate and also by
considering the van der Waals radius of these interactions.
The pair interaction noble-gas-noble-gas is stronger than
noble-gas-carbon, and the former is enough to prevent the formation of the
commensurate lattice in most cases.

For Ar atoms the minima of both the mutual interaction and the one with
the substrate occurs approximately at the same pair separation. The van
der Waals radius for both interactions are also about of the same value.
These characteristics seem enough to explain the observed behavior of the
NN distance, it increases with temperature. At density $\rho=1$ the Ne atoms
also display the above characteristic despite of different van der Waals
radius for the Ne-Ne and Ne-C interactions. Probably the observed behavior
can be explained by a very similar value of the minima of these
interactions.

Adlayers that are commensurate initially show a decrease with temperature for
the NN distance. For the Ne atoms we may attribute this to both the high
coverage this system has with respect to the substrate and a slightly
strong minimum Ne-Ne interaction. On the other hand, as the Kr atoms gain
thermal energy they become less affected by the minimum of the Kr-Kr
interaction and prefer to stay close to the minimum of the Kr-C
interaction which has also a smallest van der Waals radius. In general, it
seems the systems preference is to chose a state where the van der Waals
radius is minimum.

The third behavior we have observed, the promotion of Xe and Rn atoms at
certain temperatures to a second layer may be due to the much stronger
mutual interaction these atoms have among themselves than with the
substrate.  Although as the temperature increases the atoms would prefer
to go more near the substrate interaction minima, the mutual attraction is
so intense that the energy cost makes preferable the promotion of 
the adatoms to the
second layer.

The nearest neighbor distance is smaller than the one reported for the same 
noble gases on graphite, which is a consequence of graphene being a less 
attractive substrate. For Ne adlayers on graphite the experimental value of 
$(3.25\pm 0.02)$ \AA \ at $T=$1.5 K and $\rho=1.71$ has been 
reported\cite{tib82}, which is larger than all values we observed in the range 
$10\leqslant T \leqslant 50$ K. The experimental value\cite{dam90} for the 
nearest neighbor distance for Ar/graphite at $\rho=1$ and $T=$49.7 K is $3.97$ 
\AA, also larger than the value we observe for graphene substrates. Monte Carlo 
simulations of Ar/graphite\cite{fle06} also report larger nearest neighbor 
distances for $\rho=0.39$ and $0.71$. The system Xe/graphite exhibits a nearest 
neighbor spacing \cite{nut93} of 4.59 \AA \ at $\rho=1$ and $T=97$ K, also 
larger than our values. Moreover, the distance from the substrate also appears 
to be larger for noble gases on graphite. Simulations for Ar adlayers 
\cite{fle06} show their distance varying between $3.42$ to $3.49$ \AA \ at 
$\rho=1.14$ in the temperature range of 30 to 80 K, which is larger than our 
values.

We also computed
the average distance between the adlayer and the substrate.
As we have already discussed, it is possible 
to relate the behavior of this quantity directly to the specific heat peaks 
or the melting transition.

Future calculations will include a more detailed study of the phase
transitions, which can lead to the construction of phase diagrams. We also
intend to include more accurate potentials for the interaction between the
noble gases atoms and the graphene layer, such as the substrate-mediated
(McLachlan) potential \cite{bru10}.

This work enlarges the body of
knowledge found in the literature about 
noble gases adsorbed on carbon substrates.
To the best of our
knowledge, experimental data is not available for the noble-gas/graphene
system. We hope that this study can motivate experiments to increase our
understanding of these systems. % and to improve the simulations.

\begin{acknowledgments}
We gratefully acknowledge support from the Brazilian
agency S\~{a}o Paulo Research Foundation (FAPESP), grants 2012/24195-2 and 
2010/10072-0. 
%We thank Prof. for helpful discussions.
The calculations were performed at
CCJDR-IFGW-UNICAMP and CENAPAD-SP (project 139).
\end{acknowledgments}

\bibliography{article}% Produces the bibliography via BibTeX.

\end{document}